\documentclass[twocolumn,tighten]{aastex62}
\usepackage{amsmath}

\newcommand{\keyw}[1]{\textcolor{gray}{#1}}

\begin{document}

\title{One Solution to the Mass Budget Problem for Planet Formation: Optically Thick Disks with Dust Scattering}
\shorttitle{Optically Thick Disks}
\shortauthors{Zhu et al.}

\correspondingauthor{Zhaohuan Zhu}
\email{zhaohuan.zhu@unlv.edu}

\author[0000-0003-3616-6822]{Zhaohuan Zhu}
\affiliation{Department of Physics and Astronomy, University of Nevada, Las Vegas, 4505 S.~Maryland Pkwy, Las Vegas, NV 89154, USA}

\author[0000-0002-8537-9114]{Shangjia Zhang}
\affiliation{Department of Physics and Astronomy, University of Nevada, Las Vegas, 4505 S.~Maryland Pkwy, Las Vegas, NV 89154, USA}

\author[0000-0002-2624-3399]{Yan-Fei Jiang}
\affiliation{Kavli Institute for Theoretical Physics, University of California, Santa Barbara, CA, USA}

\author[0000-0003-4562-4119]{Akimasa Kataoka}
\affiliation{National Astronomical Observatory of Japan, Osawa 2-21-1, Mitaka, Tokyo 181-8588, Japan}

\author[0000-0002-1899-8783]{Tilman Birnstiel}
\affiliation{University Observatory, Faculty of Physics, Ludwig-Maximilians-Universit\"at M\"unchen, Scheinerstr.~1, 81679 Munich, Germany}

\author[0000-0002-7078-5910]{Cornelis P.~Dullemond}
\affiliation{Zentrum f{\"u}r Astronomie, Heidelberg University, Albert Ueberle Str.~2, 69120 Heidelberg, Germany}

\author[0000-0003-2253-2270]{Sean~M.~Andrews}
\affiliation{Harvard-Smithsonian Center for Astrophysics, 60 Garden Street, Cambridge, MA 02138, USA}

\author[0000-0001-6947-6072]{Jane Huang}
\affiliation{Harvard-Smithsonian Center for Astrophysics, 60 Garden Street, Cambridge, MA 02138, USA}

\author[0000-0002-1199-9564]{Laura M.~P{\'e}rez}
\affiliation{Departamento de Astronom{\'i}a, Universidad de Chile, Camino El Observatorio 1515, Las Condes, Santiago, Chile}

\author[0000-0003-2251-0602]{John M.~Carpenter}
\affiliation{Joint ALMA Observatory, Avenida Alonso de C{\'o}rdova 3107, Vitacura, Santiago, Chile}

\author[0000-0001-6906-9549]{Xue-Ning Bai}
\affiliation{Institute for Advanced Study and Tsinghua Center for Astrophysics, Tsinghua University, Beijing 100084, China}

\author[0000-0003-1526-7587]{David J.~Wilner}
\affiliation{Harvard-Smithsonian Center for Astrophysics, 60 Garden Street, Cambridge, MA 02138, USA}

\author{Luca Ricci}
\affiliation{Department of Physics and Astronomy, California State University Northridge, 18111 Nordhoff Street, Northridge, CA 91130, USA}

\begin{abstract}
ALMA surveys have 
suggested that the dust in Class II disks may not be enough to explain the averaged solid mass in exoplanets, under the assumption that  
the mm disk continuum emission is optically thin. This optically thin assumption seems to be supported by
recent DSHARP observations where the measured optical depths of spatially resolved disks are mostly less than one.
However, we point out that dust scattering can considerably reduce the emission from an optically thick region. If that
scattering is ignored, the optical depth will be considerably underestimated. An optically thick disk with scattering can be misidentified as
an optically thin disk. Dust scattering in more inclined disks can reduce the intensity even further, making the disk look even fainter.
The measured optical depth of $\sim$0.6 in several DSHARP disks can be naturally explained by optically thick dust with an albedo of $\sim$0.9 at 1.25 mm.
Using the DSHARP opacity, this albedo corresponds to a dust population with the maximum grain size ($s_{max}$) of 0.1-1 mm. 
For optically thick scattering disks, the measured spectral index $\alpha$ can be either larger or smaller than 2 depending on if the dust albedo increases or decreases with wavelength.  
Using the DSHARP opacity, $\alpha<2$ corresponds to $s_{max}$ of 0.03-0.3 mm. 
We describe how this optically thick scattering scenario could explain the observed scaling between submm continuum sizes
and luminosities, and might help ease the tension between the dust size constraints  from polarization and dust continuum measurements.
We suggest that a significant amount of disk mass can be hidden from ALMA observations at short millimeter wavelengths. 
For compact disks smaller than 30 au, we can easily underestimate the dust mass by more than a factor of 10. 
Longer wavelength observations (e.g. VLA or SKA) are desired to probe the dust mass in disks. 

\end{abstract}

\section{Introduction \label{sec:intro}}
The properties of protoplanetary disks (e.g. mass and size) largely determine the properties of planets born in the disks.
Radio observations play a fundamental role in studying these disks due to the presumed low dust opacity at these wavelengths.
The Atacama Large Millimeter Array (ALMA) has revolutionized protoplanetary disk studies
by providing the necessary sensitivity and spatial resolution to probe the planet forming region at several au scales for the first time \citep{brogan15, andrews16}.
Furthermore, due to ALMA's high sensitivity,  systematic surveys for a large number of young stars can be carried out efficiently. 

Previous submm/mm surveys \citep{beckwith90, aw05, andrews13}, together with recent ALMA surveys \citep{ansdell16, cieza19}, have 
suggested that
the mass of small dust ($\lesssim$cm, which ALMA is sensitive to) in Class II protostellar disks is quite low. The mean mass
is $\sim$15 $M_{\earth}$ for Lupus \citep{ansdell16}, Taurus \citep{andrews13}, and  Ophiuchus \citep{cieza19} young stars, under the assumption that the disks are optically thin at the observed wavelengths. 
This mean mass can barely explain the averaged mass of solids in exoplanets based on the exoplanet demographics derived by {\it Kepler} (e.g. \citealt{dongzhu13,chianglaughlin13,najita14}). 
This shortage of solid material in disks
becomes much more severe for lower mass stars with $\sim$ 0.4 $M_{\odot}$\citep{pascucci16}. Lower mass stars have significantly less solids in disks while they are surrounded by planets having more solids \citep{mulders15}. 
Such dust shortage leads to the speculation that dust grows efficiently when the system is
younger than the Class II phase \citep{najita14}. After the embedded Class 0 and Class I phase at an age of $\sim 10^5$ yrs \citep{tychoniec18}, 
dust may have grown larger than cm sizes, maybe even to km-sized planetesimals, so that
ALMA would not be sensitive to the majority of solids during the Class II phase.

Another solution to this mass budget problem is that Class II protostellar disks are optically thick at ALMA wavelengths
so that these disks can hide a large amount of dust mass. This idea is supported by the submm continuum size and luminosity relationship \citep{andrews10b, ricci12, tripathi17, andrews18a}.
But the optically thick disks generate too much emission and a filling factor of 0.3 in disks is needed to explain the observations.   
Recently, the high angular resolution observations from the Disk Substructures at High Angular Resolution Project (DSHARP)
\citep{andrews18b} reveal that most of the DSHARP disks have optical depths less than 1 even within the inner 20 AU \citep{huang18b}, which seems to support
the optically thin assumption. Interestingly, the maximum optical depth in some systems, such as
HD 163296 \citep{isella18}, AS 209 \citep{guzman18}, DoAr 33, and Elias 24,  seems to plateau around 0.6 (Figure 6 in \citealt{huang18b}, and \citealt{dullemond18}). 
To derive the optical depth, \cite{huang18b} fit the observed
intensity ($I_{\nu}$) at each radius $r$  using the disk midplane temperature:
\begin{equation}
I_{\nu}(r)=B_{\nu}(T_{mid}(r))(1-e^{-\tau_{\nu}(r)})\,, \label{eq:Inu}
\end{equation}
where $I_{\nu}(r)$ is the deprojected, azimuthally averaged radial intensity profile. 
The midplane temperature, which is also the temperature of mm/cm dust at the disk midplane, is estimated based on the passively heated, flared disk model
\begin{equation}
T_{mid}(r)=\left(\frac{\phi L_{*}}{8\pi r^2\sigma_{SB}}\right)^{1/4}\,,\label{eq:tmidpre}
\end{equation}
where $\sigma_{SB}$ is the Stefan-Boltzmann constant, $L_{*}$ is the stellar luminosity, and
$\phi$ is the flaring angle. \cite{huang18b} has chosen a conservative value of $\phi=0.02$, which is also used in \cite{dullemond18} and \cite{zhang18}.
Figure \ref{eq:Inu} suggests that, in an optically thin disk with the Rayleigh-Jeans approximation, $T_{mid}$ and $\tau_{\nu}$ are degenerate. If  $T_{mid}$ decreases by a factor of 2, $\tau_{\nu}$ will increase by a factor of 2. 
Thus, one might argue that the real disks actually have $\tau\gtrsim$1 instead of $\tau\sim$0.6 since 
Equation \ref{eq:tmidpre} may overestimate the disk midplane temperature by a factor of 2.
However, changing $T_{mid}$  by a factor of 2 requires $\phi$ to be changed by a factor of 16. 
For a full disk, such a large flaring angle ($\phi$) change is not supported by radiative transfer calculations \citep{dalessio98,dalessio01}.
On the other hand, we don't have direct measurements of the disk temperature and a very low temperature is still possible if 
the disk has structures which can cast shadows or the dust is highly settled at the midplane. 

{ In this paper, we point out that scattering can change the disk intensity significantly and Equation \ref{eq:Inu} needs to be modified to account for the scattering effect. 
When the disk is isothermal along the vertical direction and optically thick, 
Equation \ref{eq:Inu} reduces to $I_{\nu}=B_{\nu}$. However, this is only true for systems without scattering. 
When scattering is important, $I_{\nu}$ can be smaller than $B_{\nu}$ \citep{rybicki79}. This emission reduction can be understood intuitively using the mean free path of a photon argument.
Suppose that the single scattering albedo is
$\omega_{\nu}=\sigma_{\nu,s}/(\sigma_{\nu,a}+\sigma_{\nu,s})$ where $\sigma_{\nu,s}$ and $\sigma_{\nu,a}$
are the scattering and absorption coefficients for a photon having the frequency of $\nu$. The mean free path of a photon 
is  thus $l_{\nu}=(\sigma_{\nu,a}+\sigma_{\nu,s})^{-1}$. However, the photon needs to be scattered $(1-\omega_{\nu})^{-1}$ times before being absorbed. So after the random walk for
$(1-\omega_{\nu})^{-1}$ steps,
the mean free path for the true absorption is $l_{\nu,a}=(1-\omega_{\nu})^{-1/2}(\sigma_{\nu,a}+\sigma_{\nu,s})^{-1}$. Any photon emitted deeper 
than $l_{\nu,a}$ from the surface cannot escape. Thus, the total emission is $\sigma_{\nu,a}B_{\nu}l_{\nu,a}$ or $I_{\nu}\sim\sqrt{1-\omega_{\nu}}B_{\nu}$. Basically,
scattering reduces the depth where photons can escape. 
This smaller intensity makes an optically thick disk look optically thin.

Unfortunately, this emission reduction effect due to dust scattering has largely been ignored in previous radio intensity observations, 
despite that the rigorous derivation of this effect is presented in Appendix B of the seminal paper by \cite{miyake93}.
This omission is partly due to the assumption that scattering does not play an important role at radio wavelengths. 
On the other hand, recent ALMA polarization measurements suggest that dust scattering is crucial for explaining these
observations \citep{kataoka15}. Thus, we should also consider the effect of dust scattering on intensity measurements.  }

In \S 2, we will give the analytical solution for the isothermal disk with scattering and confirm it with numerical calculations. 
After discussing  some of the implications for the disk mass, the dust size distribution, and the spectral index in \S 3, we will conclude the paper in \S 4.

\section{Methods and Results}
In this section, we will summarize the analytical theory on radiative transfer with scattering (\S 2.1), 
and then present numerical confirmation by both direct calculations  (\S 2.2) and Monte-Carlo radiative transfer calculations (\S 2.3).

\subsection{Analytical Theory}
Consider a flat disk region with a uniform temperature of $T$. 
The intensity emitted by this region has been calculated by \cite{miyake93}.  
Here, we follow the derivation given by \cite{birnstiel18} and extend it further to very optically thick cases.

The general radiative transfer equation is
\begin{equation}
\frac{1}{c}\frac{\partial I_{\nu}}{\partial t}+ { n}\cdot\nabla I_{\nu}=-(\sigma_{\nu,a}+\sigma_{\nu,s}^{eff})I_{\nu}+j_{\nu}+\sigma_{\nu,s}^{eff}J_{\nu}\label{eq:iequationnu}
\end{equation}
where $I_{\nu}({ x},t,{ n})$ is the intensity at the position ${ x}$, time $t$ and along the direction of ${n}$. $J_{\nu}=(4\pi)^{-1}\int I_{\nu}d\Omega$ and $j_{\nu}/\sigma_{\nu,a}=B_{\nu}$, while $\sigma_{\nu,a}$ and $\sigma_{\nu,s}^{eff}$ are the absorption and effective scattering opacity at the frequency of $\nu$. This equation implicitly assumes that the scattering is isotropic. 
Since the scattering is not isotropic for the dust with sizes ($s$) $2\pi s\gg\lambda$, we use the effective scattering coefficient to approximate the anisotropic scattering effect with $\sigma_{\nu,s}^{eff}=(1-g_{\nu})\sigma_{\nu,s}$ where $g_{\nu}$ is the usual forward-scattering parameter. 
This approximation is valid for the optically thick disk \citep{ishimaru78} that is the focus of this work. 

Assuming that the disk surface follows the 1-D plane atmosphere geometry and has a time-independent radiation field, the radiative transfer equation throughout the disk is simplified to
\begin{equation}
\mu\frac{d I_{\nu}}{dz}=-(\sigma_{\nu,a}+\sigma_{\nu,s}^{eff})I_{\nu}+j_{\nu}+\sigma_{\nu,s}^{eff}J_{\nu}\,,\label{eq:iequationnu}
\end{equation}
where $\mu={\rm cos} (\theta)$ and $\theta$ is the angle between  ${ n}$ and the vertical direction (the $z$-direction). 
The 1-D plane atmosphere geometry can be justified considering that the radio emission comes from a thin disk midplane \citep{pinte16}.

If we adopt $d\tau_{\nu}=-(\sigma_{\nu,a}+\sigma_{\nu,s}^{eff})dz$, we have
\begin{equation}
\mu\frac{d I_{\nu}}{d\tau_{\nu}}=I_{\nu}-S_{\nu}\label{eq:inutau}
\end{equation}
with
\begin{equation}
S_{\nu}=(1-\omega_{\nu})B_{\nu}(T)+\omega_{\nu}J_{\nu}(\tau_{\nu})\,,\label{eq:source}
\end{equation}
where the single scattering albedo $\omega_{\nu}=\sigma_{\nu,s}^{eff}/(\sigma_{\nu,a}+\sigma_{\nu,s}^{eff})$ and $1-\omega_{\nu}=\sigma_{\nu,a}/(\sigma_{\nu,a}+\sigma_{\nu,s}^{eff})$.

With the Eddington approximation, the second moment of the radiative transfer equation becomes
\begin{equation}
\frac{1}{3}\frac{\partial^2 J}{\partial \tau_{\nu}^2}=(1-\omega_{\nu})(J_{\nu}-B_{\nu}(T))\,.
\end{equation}
If the temperature of the plane slab is a constant and there is no incoming radiation field at the upper and lower disk surface, the solution of the equation can be derived
using the two-stream approximation \citep{miyake93} as
\begin{eqnarray}
&J_{\nu}(\tau_{\nu})=B_{\nu}(T)\times\nonumber\\
&\left(1-\frac{e^{-\sqrt{3(1-\omega_{\nu})}\tau_{\nu}}+e^{\sqrt{3(1-\omega_{\nu})}(\tau_{\nu}-\tau_{\nu,d})}}{e^{-\sqrt{3(1-\omega_{\nu})}\tau_{\nu,d}}(1-\sqrt{1-\omega_{\nu}})+(\sqrt{1-\omega_{\nu}}+1)}\right)\,,\label{eq:JoB}
\end{eqnarray}
where $\tau_{\nu,d}$ and $\tau_{\nu}$ are the total and variable optical depth in the vertical direction. 

With $J_{\nu}$ known, we can integrate Equation \ref{eq:inutau} throughout the disk to derive the emergent intensity ($I_{\nu}^{out}$). 
Based on the Eddington-Barbier approximation, the solution \citep{birnstiel18} is
\begin{equation}
I_{\nu}^{out}=(1-e^{-\tau_{\nu,d}/\mu})S_{\nu}(\tau_{\nu}=2\mu/3)\label{eq:Inuout}
\end{equation}
with $S_{\nu}(\tau_{\nu})$ given in Equation \ref{eq:source} and $J_{\nu}(\tau_{\nu})$ in Equation \ref{eq:JoB}.
If $\tau_{\nu,d}<4\mu/3$, $\tau_{\nu}$ in $S_{\nu}$ is chosen as $\tau_{\nu,d}/2$. 

{ With $I_{\nu}^{out}$, we can define its deviation from the blackbody radiation using
\begin{equation}
\chi\equiv\frac{I_{\nu}^{out}}{B_{\nu}}\,.\label{eq:chidef}
\end{equation}
As alluded to in the introduction, $\chi<1$ can be due to either emission from the optically thin region or dust scattering in the optically thick region.
}

{ If we choose $\tau_{\nu}=2\mu\tau_{\nu,d}/(3\tau_{\nu,d}+1)$ in Equation \ref{eq:Inuout} to 
approximate both the optically thick and thin cases, Equations \ref{eq:Inuout} and \ref{eq:chidef} can be written out explicitly as
\begin{eqnarray}
&\chi\equiv\frac{I_{\nu}^{out}}{B_{\nu}}=(1-e^{-\tau_{\nu,d}/\mu})\times \nonumber\\
&\left(1-\omega_{\nu}\frac{e^{-\sqrt{3(1-\omega_{\nu})}\tau_{\nu}}+e^{\sqrt{3(1-\omega_{\nu})}(\tau_{\nu}-\tau_{\nu,d})}}{e^{-\sqrt{3(1-\omega_{\nu})}\tau_{\nu,d}}(1-\sqrt{1-\omega_{\nu}})+(\sqrt{1-\omega_{\nu}}+1)}\right)\nonumber\\
& {\rm with}\,\,\,\tau_{\nu}=2\mu\tau_{\nu,d}/(3\tau_{\nu,d}+1)\,.\label{eq:chisimple}
\end{eqnarray}

}


For the optically thin region, Equation \ref{eq:Inuout} or \ref{eq:chisimple} reduces to $I_{\nu}^{out}\rightarrow (1-\omega_{\nu})\tau_{\nu,d}B_{\nu}/\mu$.
The quantity of $(1-\omega_{\nu})\tau_{\nu,d}$ is basically the disk optical depth calculated with the absorption coefficient or $\tau_{\nu,d}^{abs}$. 
Thus, when the disk is optically thin ($\tau_{\nu,d}<1$), the emergent intensity reflects only the absorption opacity.

On the other hand, when the disk is optically thick with $(1-\omega_{\nu})\tau_{\nu,d}\gg1$, the emission for a disk
with scattering is smaller than a disk without scattering (the black body radiation) by a factor of 
\begin{equation}
\chi\equiv\frac{I_{\nu}^{out}}{B_{\nu}}=1-\frac{\omega_{\nu}}{(\sqrt{1-\omega_{\nu}}+1)e^{\sqrt{3(1-\omega_{\nu})}\cdot 2\mu/3}}\,. \label{eq:Inuoutthick}
\end{equation}
Since $\chi<1$ if $\omega_{\nu}>0$, the optically thick scattering disk looks fainter than the blackbody radiation calculated using the same disk temperature.

If we expand the exponent in Equation \ref{eq:Inuoutthick} with the Taylor series, we can simplify the equation further to
\begin{equation}
   \chi=\frac{\sqrt{3}+2\mu}{2\mu+\frac{\sqrt{3}}{\sqrt{1-\omega_{\nu}}}}\,. \label{eq:Inuoutthick2}
\end{equation}
{ In the extreme case with $\omega_{\nu}\rightarrow$1, Equation \ref{eq:Inuoutthick2} becomes $\chi=(1+1.15 \mu)\sqrt{1-\omega_{\nu}}\sim\sqrt{1-\omega_{\nu}}$, which 
is similar to the result based on the mean free path argument in the introduction.} We have verified that Equation \ref{eq:Inuoutthick2} only deviates from Equation \ref{eq:Inuoutthick} by less than 5\% over the whole parameter space.
This enables us to solve $\omega_{\nu}$ analytically using $\chi$, as
\begin{equation}
\omega_{\nu}=1-\left(\frac{1}{\chi}+\left(\frac{1}{\chi}-1\right)\frac{2\mu}{\sqrt{3}}\right)^{-2}\,.\label{eq:omeganucal}
\end{equation}

If we just apply  Equation \ref{eq:Inu} to calculating the disk optical depth using the emergent intensity from an optically thick disk (Equation \ref{eq:Inuoutthick}), we will derive
an optical depth of
\begin{equation}
\tau_{obs}=-{\rm ln} (1-\chi)\,,\label{eq:tauobs}
\end{equation}
even if the disk is very optically thick. Thus, another explanation for $\tau_{obs}=0.6$ in the DSHARP disks is that these disks are actually very optically thick (e.g. $\tau=10^4$) but with
$\chi=0.45$ due to dust scattering (by plugging $\tau_{obs}=0.6$ into Equation \ref{eq:tauobs}) . With $\mu$=1 or 0.5, $\chi$=0.45 corresponds to $\omega_{\nu}$=0.93 or 0.89 respectively (Equation \ref{eq:omeganucal}), suggesting that 
the dust in these disks is highly reflective.

\begin{figure}[t!]
\includegraphics[width=3.5in]{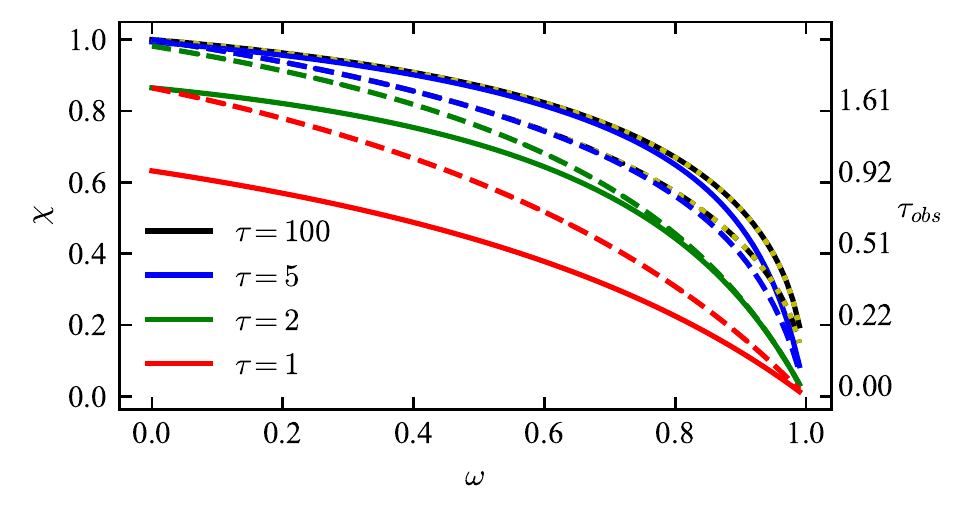}
\figcaption{The intensity reduction factor $\chi$ with respect to $\omega_{\nu}$ for disks with different optical depths using Equation \ref{eq:Inuout}. The solid curves
are derived with $\mu$=1 ($i=0^{\circ}$, face-on) while the dashed curves are derived with $\mu$=0.5 ($i=60^{\circ}$).  The yellow dotted curves, which are basically on top of the black curves, are
derived with the asymptotic optically thick limit (Equation \ref{eq:Inuoutthick}). The corresponding $\tau_{obs}$ based on
Equation \ref{eq:tauobs} is shown on the right axis. \label{fig:omega}}
\end{figure}

\begin{figure}[t!]
\includegraphics[width=3.5in]{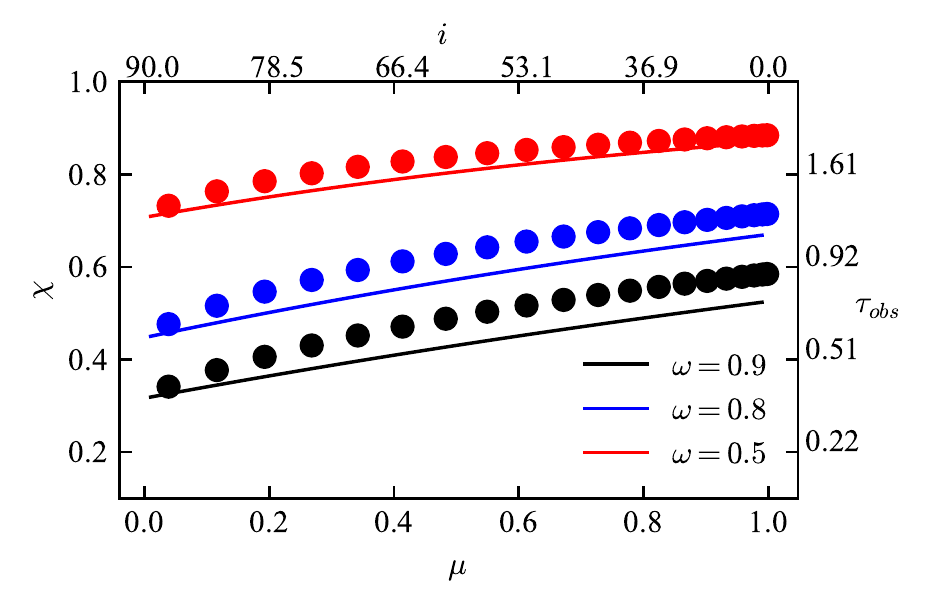}
\figcaption{$\chi$ with respect to the inclination ($\mu$=cos $i$ for the bottom axis and $i$ for the top axis) for different albedos ($\omega$). The solid curves are from the analytical estimate in the optically thick limit (Equation \ref{eq:Inuoutthick})
while the dots are from direct numerical simulations. 
The corresponding $\tau_{obs}$ is shown on the right axis. \label{fig:mu}}
\end{figure}

Figure \ref{fig:omega} shows how $\chi$ changes with $\omega$ using Equation \ref{eq:Inuout}.
As expected, the intensity drops when the disk becomes more optically thin. On the other hand,
even if the disk is optically thick, stronger scattering can also lead to a smaller intensity. 
In the optically thick limit ($\tau=100$), Equation  \ref{eq:Inuoutthick} (yellow curves) agrees with the full solution (Equation \ref{eq:Inuout}) very well.
For the marginally optically-thick or optically-thin disks (the green and red curves), the intensity increases when the disk is more inclined (dashed curves). This is simply because our line of sight
passes through more column ($1/\mu$ factor) when the disk is inclined. 
On the other hand, for very optically-thick disks ($\tau\gtrsim 5$), the intensity decreases when the disk is more inclined due to dust scattering. Thus, inclined optically-thick disks look
even fainter than face-on disks. The change in $\chi$ with respect to the disk inclination in the optically thick limit is shown in Figure \ref{fig:mu}.

\begin{figure}[t!]
\includegraphics[width=3.5in]{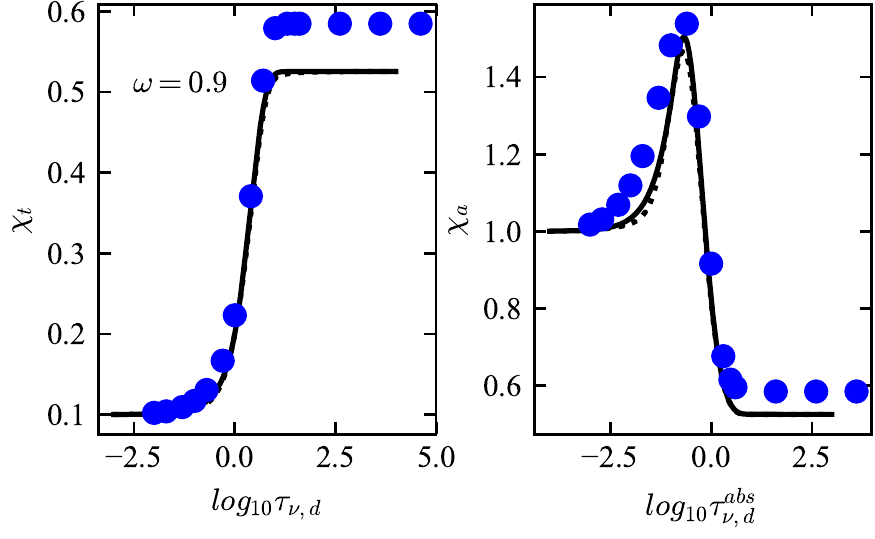}
\figcaption{$\chi_t$ with respect to the total optical depth of disks with different optical depths (the left panel), 
and $\chi_a$ with respect to the absorption optical depth of these disks (the right panel). The solid curves are derived from Equation \ref{eq:Inuout}, while
the dotted curves are from the approximated solution (Equation \ref{eq:chisimple}).
The albedo $\omega$ is set to be 0.9. The blue dots are from
direct numerical simulations. \label{fig:tau}}
\end{figure}

\begin{figure}[t!]
\includegraphics[width=3.3in]{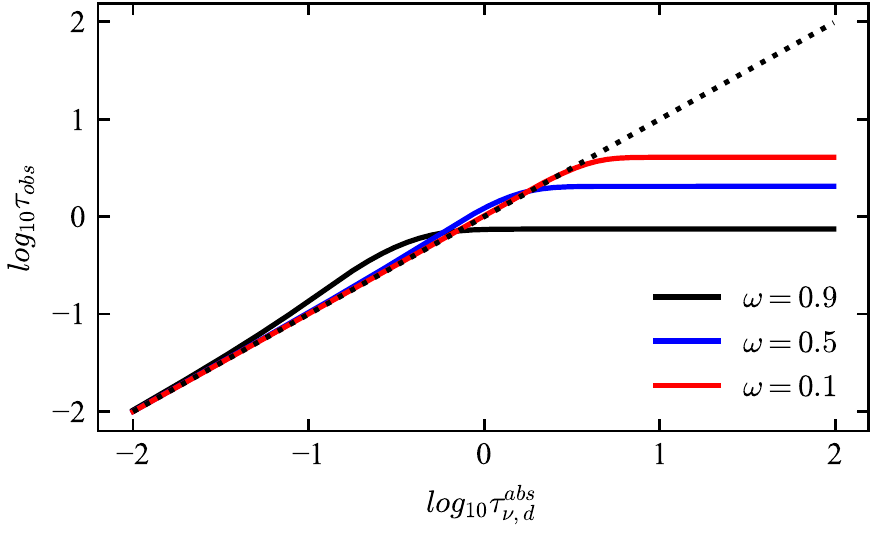}
\figcaption{The observed disk optical depth with respect to the true disk absorption optical depth. The dotted line shows $\tau_{obs}=\tau_{\nu,d}^{abs}$.\label{fig:tauobs}}
\end{figure}

To study the effect of dust scattering, we can also compare the intensity from disks having strong scattering with the intensity from disks having zero albedo.  
We thus define
\begin{eqnarray}
\chi_{t}&=&\frac{I_{\nu}^{out}}{B_{\nu}(1-e^{-\tau_{\nu,d}})}\\
\chi_{a}&=&\frac{I_{\nu}^{out}}{B_{\nu}(1-e^{-\tau_{\nu,d}^{abs}})}\,.
\end{eqnarray}
These comparisons assume that $\mu$=1. 
Without dust scattering, we have $\chi_t=\chi_a=1$. 
With dust scattering, we have $\chi_t=(1-\omega_{\nu})$ and $\chi_a=1$ in the optically thin limit. 
Figure \ref{fig:tau} shows how both $\chi_{t}$ and $\chi_{a}$ change with $\tau_{\nu,d}$ and $\tau_{\nu,d}^{abs}$ if $\omega$=0.9. When $\tau_{\nu,d}<1$, 
$\chi_t=(1-\omega_{\nu})$ is a good approximation. When $\tau_{\nu,d}^{abs}>1$, Equation \ref{eq:Inuoutthick} is a good approximation.  
Figure \ref{fig:tau} also shows an interesting phenomenon that dust scattering is not always reducing the intensity.
When $\tau_{\nu,d}^{abs}\sim 1$, the disk with scattering is actually brighter than the disk without scattering, as long as these two disks
have the same absorption optical depth. Due to this complex phenomenon at $\tau_{\nu,d}^{abs}\sim1$, when we call a disk ``optically thin" in the rest of the paper we refer
to $\tau_{\nu,d}<1$, and when we call a disk ``optically thick" we refer
to $\tau_{\nu,d}^{abs}>1$, unless otherwise stated.  

{ We also want to study how ``wrong'' the derived optical depth can be if we use the traditional method (Equation \ref{eq:Inu}) to measure the optical depth of a disk having scattering.
We first calculate the intensity emitted by the disk with the absorption optical depth of $\tau_{\nu,d}^{abs}$ and the scattering albedo ($\omega_{\nu}$) using Equation \ref{eq:Inuout}.
Then, we use Equation \ref{eq:Inu} to derive the observed optical depth ($\tau_{obs}$), assuming that we know the actual disk temperature.}
The relationship between the observed disk optical depth and the true disk optical depth is shown in Figure \ref{fig:tauobs}
for different disk albedos. 
Clearly, even if $\omega$ is only 0.1, an extremely optically thick disk can be misidentified as a disk with 
the optical depth of order  unity. 

\subsection{Direct Numerical Simulations}
To validate the approximations used in the above section (e.g. the Eddington, two-stream, and Eddington-Barbier approximations), we have carried out direct radiative transfer calculations using the radiation module
in Athena++ \citep{jiang14}. It solves the radiative transfer equation explicitly with the method of short characteristics.  Here we only solve the radiative transfer equation without evolving the hydrodynamics. 

We set up a plane-parallel atmosphere with a density profile of
\begin{equation}
\rho=\rho_{0} e^{-(z^2-z_{min}^2)/2H^2}\,,\label{eq:simpleatmosphere1}\\
\end{equation}
to represent the disk vertical density structure,
where $\rho_{0}=1$ and  $H=0.05$ in the code unit. 
The simulation domain extends from { the midplane} at $z_{min}=0$ to $z_{max}=0.35$ with 256 uniform grid cells.
For the radiation field, the reflecting boundary condition ({ which flips the $z$ direction of the intensity rays}) has been adopted at the disk midplane $z_{min}$, { considering that the disk is symmetric with respect to the midplane}. 
The vacuum boundary condition has been adopted at $z_{max}$ to simulate the outflowing radiation field. 
We vary the opacity to control the optical depth of the disk but keep $\omega$=0.9 for all the simulations. We solve the radiative transfer
equation along 40 different angles.  

\begin{figure}[t!]
\includegraphics[width=3.3in]{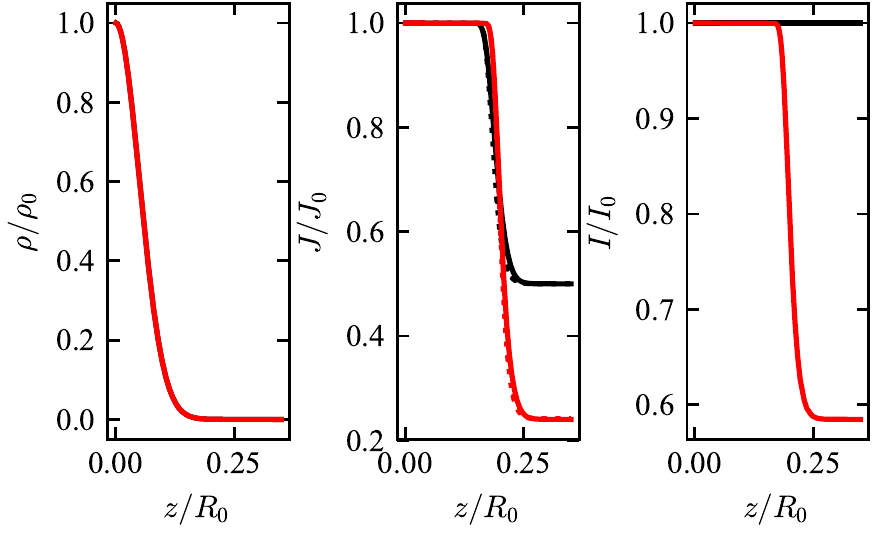}
\figcaption{The profiles of the disk density, J, and I from our fiducial simulation. The black solid curves are the case with $\omega=0$ and $\tau_{\nu,d}^{abs}$=4,000,
while the red solid curves are the case with the $\omega=0.9$ isotropic scattering and the same $\tau_{\nu,d}^{abs}$.  The dotted curves in the middle panel 
are $J$ calculated from the analytical theory. \label{fig:J}}
\end{figure}

Figure \ref{fig:J} shows the density structure and the radiation field from our fiducial simulation. The rightmost panel shows the intensity in the direction that is perpendicular to the disk surface. The red curves are from disks with
$\omega$=0.9 and $\tau_{\nu,d}=40,000$, while the black curves are from disks without scattering and $\tau_{\nu,d}=4,000$. Both
disks have $\tau_{\nu,d}^{abs}=4,000$. The dotted curves in the middle panel are calculated using the analytical theory (Equation \ref{eq:JoB}).
As clearly shown, the analytical theory reproduces the radiation field in the simulation very well, and scattering decreases the mean field $J$ at the disk surface.  
The $J$ panel also demonstrates why scattering can reduce the emergent intensity in the optically thick limit.
Scattering couples the emergent intensity with $J$ (Equations \ref{eq:source} and \ref{eq:Inuout})  that deceases at the disk surface.
For the intensity coming out of the disk at other angles and the intensity from disks with different optical depths, the simulation data are plotted 
against the analytical theory in  Figures \ref{fig:mu} and \ref{fig:tau}. It seems that the analytical theory can explain the simulation results
reasonably well, although it can underpredict $\chi$ by up to 15\% in the optically thick limit. This is probably due to the approximations 
used in the analytical calculations. 

\subsection{Monte-Carlo Radiative Transfer (MCRT) Calculations }

\begin{figure*}[t!]
\includegraphics[width=6.5in]{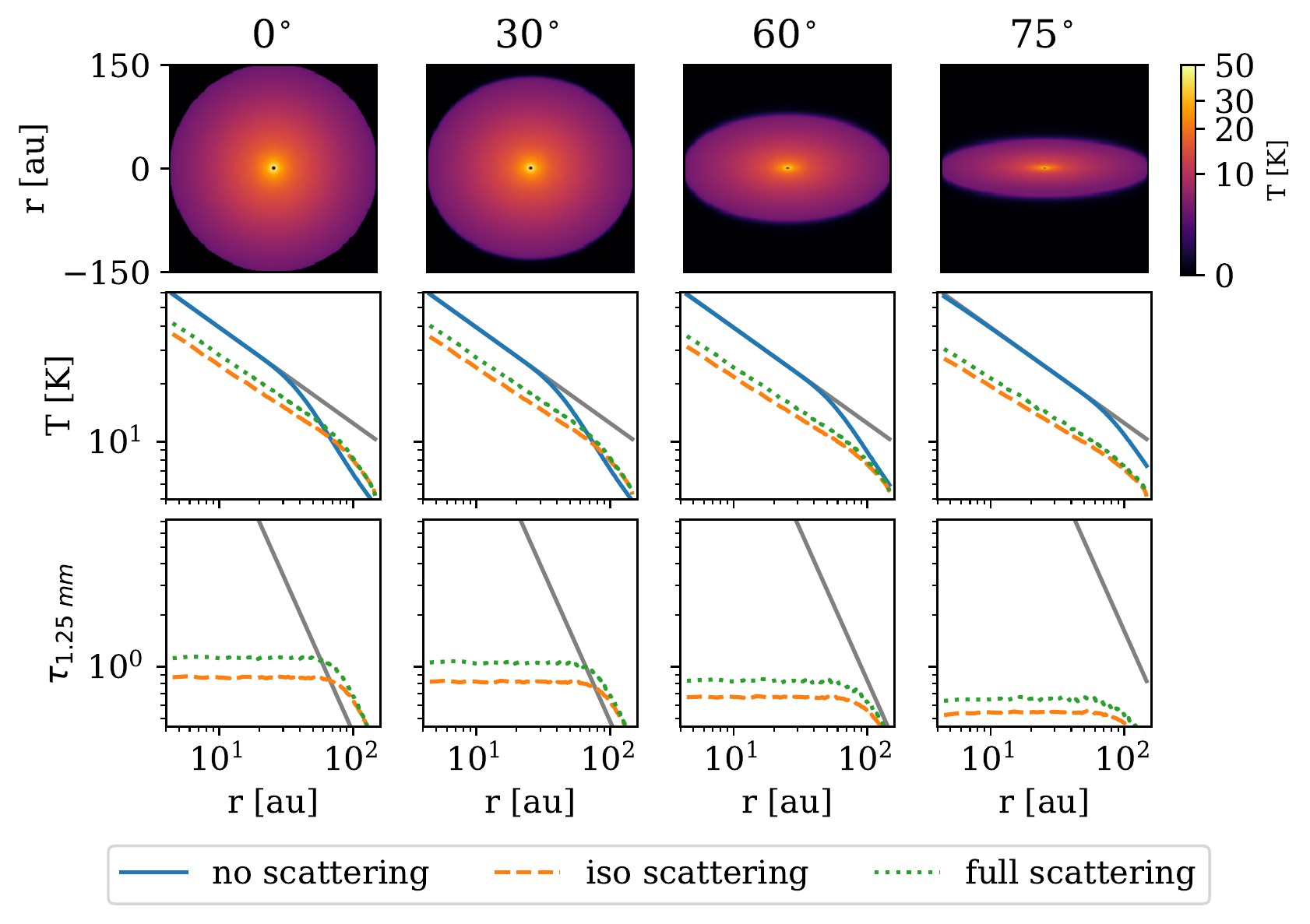}
\figcaption{The  MCRT calculations for a Q=1 protoplanetary disk viewed at different angles at 1.25 mm.  The top panels show the disk intensity for full scattering treatment. 
The middle panels show the input disk temperature (grey lines), the brightness temperature for simulations without dust scattering (blue curves),
the brightness temperature for simulations with isotropic scattering (orange dashed curves) and anisotropic scattering (green dotted curves).
The bottom panels show the measured optical depth based on the brightness temperature using Equation \ref{eq:Inu}. The grey lines are $\tau$ estimated by $\kappa_{\nu}^{abs}\Sigma_d(r)/\mu$. Clearly, even if the disks are highly optically thick, the measured optical
depths for disks with scattering are $\lesssim$1. More inclined scattering disks also have smaller measured optical depths.
\label{fig:scat}}
\end{figure*}

Our calculations above assume that the scattering is isotropic.  
To test anisotropic scattering, we have carried out MCRT calculations using RADMC-3D 
\footnote{RADMC-3D is an open code of radiative transfer calculations. The code is
available online:
http://www.ita.uni-heidelberg.de/~dullemond/software/radmc-3d/.}. We set up the disk with Toomre $Q=1$, which is the most massive disk possible.
The disk temperature is prescribed as in Equation \ref{eq:tmidpre}, using $L_*$ = $L_\odot$. The disk is locally isothermal along the z direction at a given $r$. We also assume $M_*$ = $M_\odot$. The disk scale height $H(r)/r$ is calculated from $H(r)/r=c_{s}/v_{\phi}$ where $c_{s}^2=RT/\mu$ and $\mu$ = 2.35. To keep $Q$ = 1 throughout the disk, the gas surface density
\begin{equation}
\Sigma_g(r) = 330 \left(\frac{r}{20 \,{\rm au}}\right)^{-1.75} {\rm g \,cm}^{-2}. \label{eq:Q1}
\end{equation}
To be in hydrostatic equilibrium, 
the disk has a Gaussian density profile along the $z$ direction,
\begin{equation}
    \rho_g(r, z) = \frac{\Sigma_g(r)}{\sqrt{2\pi}H(r)} \exp{\Big(-\frac{z^2}{2H(r)^2}\Big)}.
\end{equation}
The disk is truncated at $r_{in}$ = 3 au and $r_{out}$ = 150 au in the radial direction. 
At 20 au, $H(r)/r$ = 0.047 and T = 28 K. The dust density is set to be 1/100 of the gas density and the dust scale height is 1/5 of the gas scale height. The DSHARP opacity (with water ice) is adopted, with $s_{min}=0.1$ $\mu$m, $s_{max}$ = 1 mm and $n(s) \sim s^{-3.5}$. The composition and optical constants are the same as in Table 1 and Figure 2 of \cite{birnstiel18} and can be obtained by \texttt{dsharp\_opac} \footnote{https://github.com/birnstiel/dsharp\_opac}. For full anisotropic scattering, the M\"ueller matrices are calculated using Mie theory, specifically Bohren-Huffman program \citep{bohren83}. For isotropic scattering calculations, the opacity that is normalized to the dust density is  $\kappa_{abs}|_d$ = 2.1 cm$^2$ g$^{-1}$ and $\kappa_{sca}|_d$ = 19.5 cm$^2$ g$^{-1}$  at 1.25 mm (so $\omega$=0.9). 
To compare with the rest of the paper where the opacity is normalized to the gas density, we can derive the gas density normalized opacity $\kappa_{abs}$ = 0.021 cm$^2$ g$^{-1}$ and $\kappa_{sca}$ = 0.195 cm$^2$ g$^{-1}$. 
Here, $\rho_d\kappa_{abs,d}$  or 100$\rho_d\kappa_{abs}$ are basically $\sigma_{\nu,a}$ in Section 2.1. For the face-on disk, $\tau \sim$ 7 at 20 au.
For every disk inclination, we have run three simulations: one without scattering,
one with isotropic scattering, and one with full anisotropic scattering treatment. All these three simulations have the same absorption opacity.
$5\times 10^{8}$ photon packages have been used. The resolution in the radial, poloidal, and azimuthal directions are 512, 2048, and 32 cells respectively. The cell size in the radial direction is uniform in logarithmic space, while the cell size in the poloidal and azimuthal directions are uniform in linear space from 
 0 to 50$^{\circ}$, and from 0 to 2$\pi$ respectively. The reflecting boundary condition is used at the disk midplane. Such high resolution in the poloidal direction is crucial for  treating the scattering process properly. 

The results are shown in Figure \ref{fig:scat}.  The top panels show the
2-D intensity maps at 1.25 mm for the disks with anisotropic scattering. Intensity maps from MCRT calculations without scattering and with isotropic scattering are also generated.
We cut through the horizontal major axis in the images to derive the 1-D profiles which are shown in the middle and bottom panels. 
The middle panels show the  1-D profiles of the brightness temperature that is converted from the measured intensity. When the disk is face on, $\chi\sim0.6$ which is consistent with
our analytical estimate using $\omega=0.9$.
 The bottom panels show the derived optical depths using Equation \ref{eq:Inu}. Different colored curves show disks with different scattering treatments. We do not show the measured optical depth for no-scattering cases, since, when the optical depth becomes very large (e.g. $>$10), Equation \ref{eq:Inu}  cannot provide an accurate estimate of the optical depth.  Compared with isotropic scattering, full anisotropic scattering treatment does not
change the results qualitatively.
Clearly, if the disks are highly optically thick but have scattering, the measured optical
depths are $\lesssim$1 using Equation \ref{eq:Inu}. If the scattering disk is more inclined (to the right panels), the measured brightness temperature is smaller and thus the derived optical depth becomes smaller.

Figure \ref{fig:scat} also suggests that even a Q=1 disk will become optically thin beyond 50 au, where the brightness temperature decreases much faster than the midplane temperature. Thus, if the measured $\tau$ at the outer disk is small, the disk is probably truly optically thin instead of optically thick with strong scattering (the possible tests using the spectral index are presented in Section 3.3). 

\section{Discussion}

\subsection{Dust Mass in Disks and Future Observations}

\begin{figure}[t!]
\includegraphics[width=3.3 in]{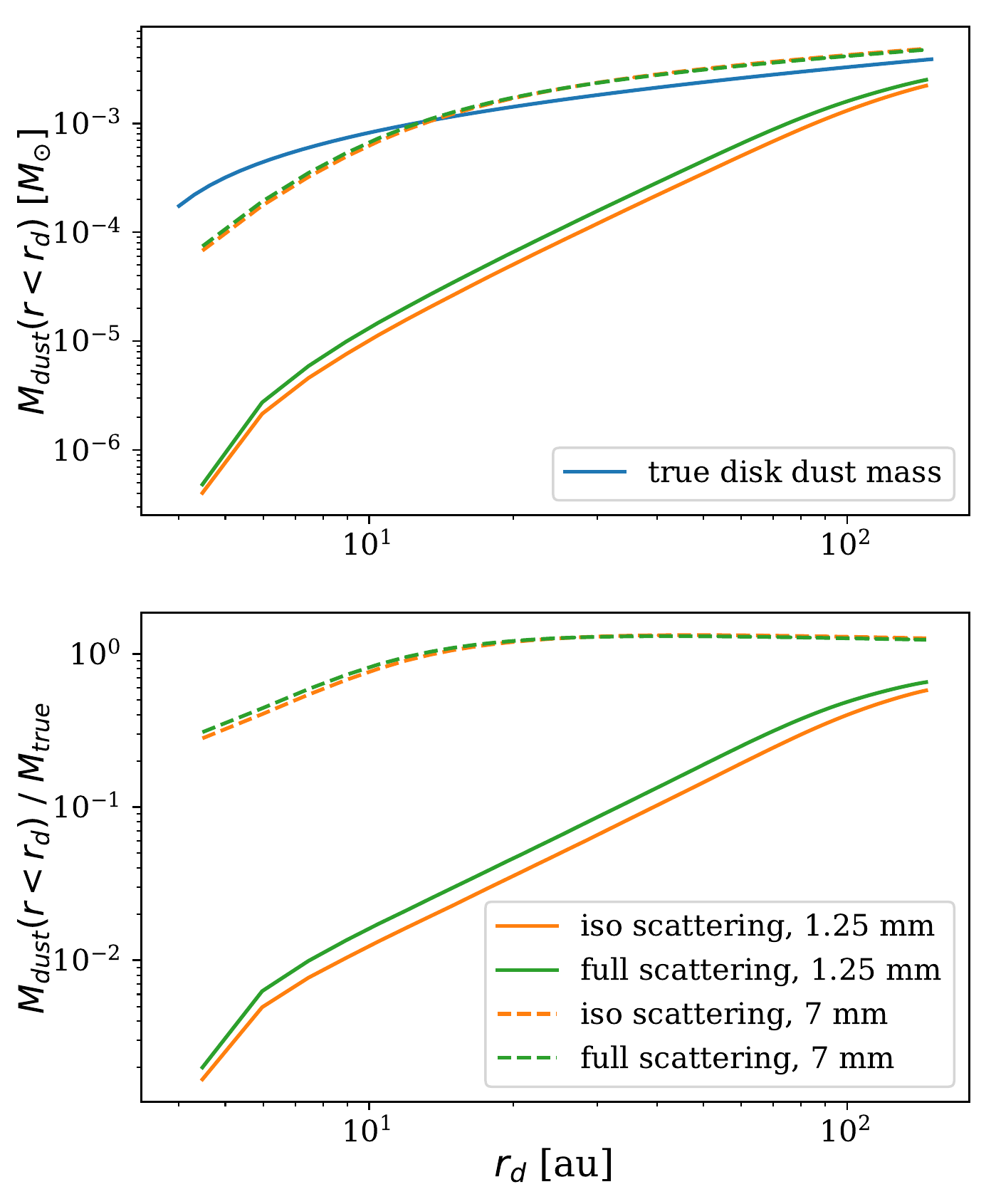}
\vspace{-1 mm}
\figcaption{{ The dust mass within $r_d$ (the upper panel) and the ratio between the measured mass within $r_d$ and the true mass within $r_d$(the lower panel)
as a function of $r_d$. Compared with the real disk mass labeled as the blue curve,
the orange and green curves are measured disk mass based on Equation \ref{eq:Inu}, where the intensity 
is measured from MCRT calculations with dust scattering (Figure \ref{fig:scat}). Two different treatments for dust scattering (two different colors) have been considered in MCRT calculations.
The solid curves are for the ALMA 1.25 mm observations while the dashed curves are for the VLA 7 mm observations.}
 \label{fig:mass}}
\end{figure}

Our proposed optically thick disk with  scattering scenario implies that protoplanetary disks can potentially hide a large amount of solids in disks (within 50 au for the Q$\sim$1 disk) from ALMA. 
Previous submm surveys may underestimate the solids in disks significantly. Class II disks may still have a significant amount of dust to form planetesimals or planets later. Such disks with more solids
are more consistent with the amount of solid mass in exoplanets. Similar to our simulations, \cite{evans17} have carried out MCRT simulations for gravitationally unstable disks and conclude that such disks can hide a factor of 3-30 dust mass to ALMA observations, even if dust scattering has not been included in these calculations.  Furthermore, the massive disks are more consistent with the fact that Class II disks are still accreting at moderate rates \citep{hartmann06}, implying
at least $10^{-8}M_{\odot}/yr\times 10^{6} yrs\sim0.01 M_{\odot}$ gas in disks. 

To study the true amount of dust in protoplanetary disks, we may need to go to longer wavelengths. VLA, ngVLA \citep{murphy18} or SKA \citep{testi15} will be quite powerful in this regard, although these long wavelength observations will get significant contamination from free-free emission from the star or jet. Assuming that  the absorption opacity changes with the frequency as $\kappa_{abs}= 0.021 (\nu/240 {\rm GHz})^{\beta}$ cm$^2$ g$^{-1}$,
$\kappa_{abs}$ is $3.6\times10^{-5}$ cm$^2$ g$^{-1}$ at 10 GHz with $\beta=2$. Using Equation \ref{eq:Q1}, we can derive that the Q=1 disk will be optically thin ($\tau_{\nu,d}^{abs}<1$) beyond 2 au at 10 GHz. On the other hand, if the dust is big (Section 3.3) and $\beta=1$, $\kappa_{abs}$ is $8.8\times10^{-4}$ cm$^2$ g$^{-1}$ at 10 GHz, and the Q=1 disk will be optically thin beyond 10 au at 10 GHz. 

{ To illustrate how much dust mass can be hidden from ALMA observations at short millimeter wavelengths in our Q=1 disk with dust scattering, we plot the measured and real dust mass in Figure \ref{fig:mass}. The disk intensity
is from the MCRT calculations in Section 2.3. Besides the 1.25 mm band in Section 2.3, we also carry out MCRT calculations at 7 mm band. Since the input disk
is as massive as a disk can get (Q=1),  Figure \ref{fig:mass} shows the maximum amount of dust mass that can be hidden from observations. When the disk is large (e.g. 100 au),
most of dust mass is at the outer disk which is optically thin at 1.25 mm, so that 1.25 mm observations will only underestimate the disk mass by a factor of $\sim$2 for these extended disks. 
When the disk is compact ($<$ 30 au), ALMA 1.25 mm observations can easily underestimate the real dust mass by a factor of 10. ALMA protoplanetary disk surveys suggest that
most disks are actually compact disks (Figure 2 in \citealt{ansdell16}), which can be due to dust radial drift. Thus, these surveys may underestimate the  dust mass significantly for the whole sample.

On the other hand, for VLA 7 mm observations,
the disk is optically thin beyond several au.  Thus, VLA observations provide a much more accurate mass estimate.  We notice that the estimated dust mass
is slightly larger than the real dust mass when the disk is optically thin, which is due to that the brightness temperature is slightly higher with scattering included (Figure \ref{fig:scat}).
We suspect that this is due to the intensity enhancing effect in Figure \ref{fig:tau}.
Recent VLA observations by \cite{tychoniec18} suggest that Class 0/I
objects are much more massive than Class II objects. However, we caution that this large difference may be due to  the fact that observations for Class 0/I and II disks are carried out at different bands.
Similar VLA surveys for Class II disks are desired to probe the real dust mass in these disks. 
}

\subsection{Constraining Dust Properties Using $\chi$}

\begin{figure*}[t!]
\includegraphics[width=5.8 in]{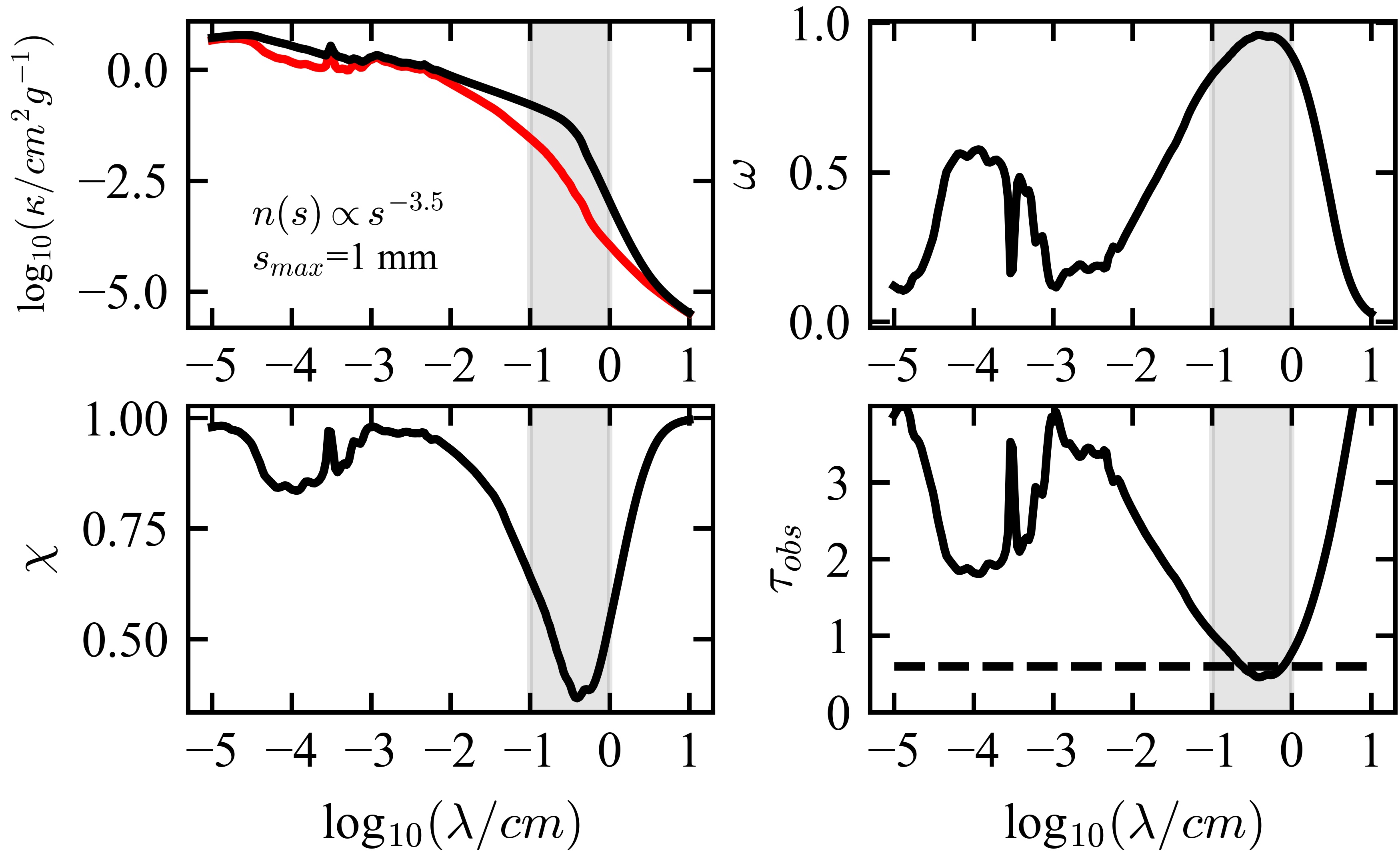}
\vspace{-1 mm}
\figcaption{The dust opacity (the upper left panel) and albedo (the upper right panel)  for a population of dust 
with the $n(s)\propto s^{-3.5}$ size distribution, $s_{min}=0.1\mu$m  and $s_{max}=1$ mm. { The dust composition and optical constants
are given in Table 1 of \cite{birnstiel18} (DSHARP opacity).}
The red and black curves in the upper left panel 
are the absorption and total dust opacity respectively. 
The calculated $\chi$ and $\tau_{obs}$ (Equation \ref{eq:tauobs}) for optically thick disks are shown in the bottom panels. The x-axis in all the panels
are the observation wavelength. The horizontal dashed line in the lower right panel labels where $\tau_{obs}=0.6$.
 \label{fig:lambda}}
\end{figure*}

{ Previously, it is proposed that we can measure the disk temperature by observing the dust continuum at higher frequencies (e.g. ALMA Bands 8, 9),
since the disk is optically thick at those bands (e.g. \citealt{kim19}). We point out that if the disk has scattering we can not use this method to measure the disk
temperature since the intensity from thermal radiation is affected by the scattering. Furthermore, we can not use ALMA to measure the disk mass accurately if the disk is 
optically thick. }

On the other hand, the simple relationships between the reduced emission ($\chi$)
and albedo (Equations \ref{eq:Inuoutthick}, \ref{eq:Inuoutthick2}, and \ref{eq:omeganucal})
 provide an unique opportunity to study dust properties at the $\tau\sim1$ surface. These relationships are independent of any particular
 disk model (e.g. whether turbulent or not), which is why they are so powerful to constrain the dust properties. 
We can measure $\chi$ if we know the disk temperature and use $\chi$ to constrain the dust albedo directly.
If we have multi-band observations, we can also use the spectral index to constrain the change in albedo (Section 3.3).

Although optically thick disks with $\omega\sim0.9$ can explain $\tau_{obs}\sim$0.6 in DSHARP observations, it is crucial
to understand whether dust in protoplanetary disks can have such a high albedo. 
Assuming that the dust
follows the $n(s)\propto s^{-3.5}$ size distribution with $s_{max}=1$ mm,
we use the DSHARP 
opacity \citep{birnstiel18} to calculate the dust opacity and albedo at different wavelengths, as shown in Figure \ref{fig:lambda}. 
Clearly, the albedo can be as high as 0.9 for radio observations at mm to cm. Dust scattering is most efficient when 2$\pi s\sim \lambda$.
With $s_{max}=1$ mm, the strongest scattering ($\omega\sim$0.96) occurs at 4 mm. With these albedos, we can calculate $\chi$
and $\tau_{obs}$ at different wavelengths assuming that the disk is  optically thick (with Equations \ref{eq:Inuoutthick} and \ref{eq:tauobs}), as shown in the bottom panels of Figure \ref{fig:lambda}.  
Clearly, for mm-cm observations (shaded region), $\chi$ is less than 0.7 and $\tau_{obs}$ is less than 1 with our assumed dust population. 
Thus, this assumed dust population can naturally explain the $\tau_{obs}<1$ in DSHARP observations.

On the other hand, we caution that the DSHARP opacity adopted here is for compact spheres without porosity. And the mixture of different compositions 
are handled with Bruggeman rule. Changing composition or porosity can change the opacity and albedo dramatically, as alluded in \cite{birnstiel18}.
{ For example, the dust becomes more reflective with more water ice. Porosity reduces the resonant opacity features, making the opacity curve smoother.  Carbonaceous materials
have a large effect on the dust opacity. The maximum absorption feature can be shifted to much longer wavelengths than $2\pi s_{max}$ with some choices of carbonaceous materials. 
On the other hand,  the peak of albedo is still at $2\pi s_{max}$ with different choices of carbonaceous materials \citep{birnstiel18}, making both $\chi$ and $\alpha$ measurements (Section 3.3)
less sensitive to the choices of carbonaceous materials.} More detailed calculations
exploring different dust compositions, porosities, and size distributions are needed in future.

\begin{figure*}[t!]
\includegraphics[width=7.4in]{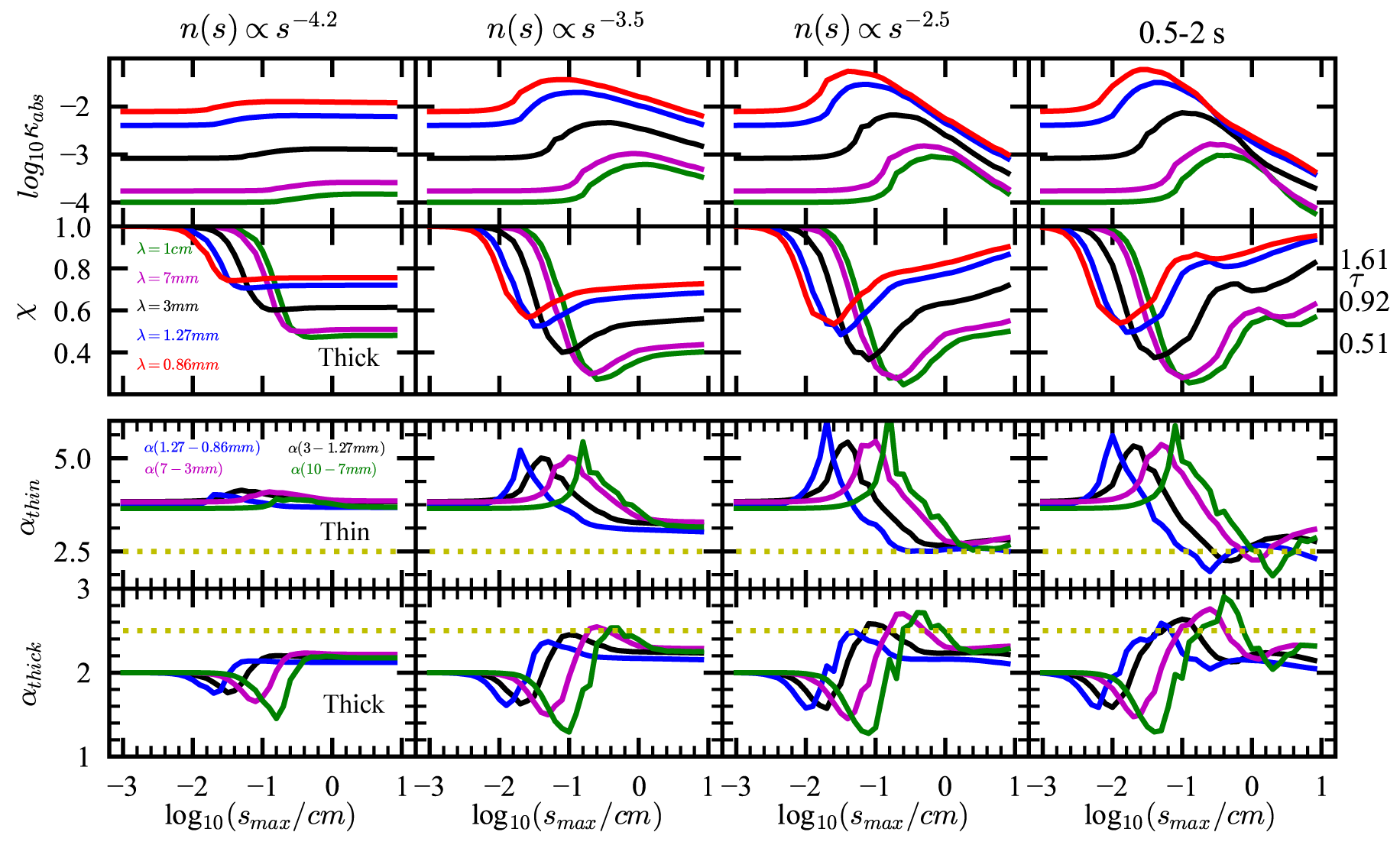}
\vspace{-4 mm}
\figcaption{$\kappa_{abs}$, $\chi$ and the spectral index $\alpha$ for different dust populations with different distributions (from left to right panels) and different
$s_{max}$ (the x-axis). The panels in the first row show the absorption opacity, while the second row shows
$\chi$ in the optically thick limit which is directly related to albedo.
The third row shows $\alpha$ in the optically thin limit (thus denoted as $\alpha_{thin}$) while the bottom row is $\alpha$ in the optically thick limit ($\alpha_{thick}$). 
The rightmost panels are for particles with a narrow size bin distribution (from 0.5 to 2 $s$ with $n(s)\propto s^{-4}$).
In the upper two rows, $\kappa_{abs}$ and $\chi$ for observations at different bands are plotted as different colors, while in the bottom two rows, $\alpha$
from different band combinations is plotted as different colors. The dotted lines in the bottom two rows label $\alpha=2.5$ for comparison. 
$s_{min}=0.1\mu$m in all these calculations. {(Section 3.3 for details)}
\label{fig:smcombined}}
\end{figure*}

\begin{figure*}[t!]
\includegraphics[width=7.2in]{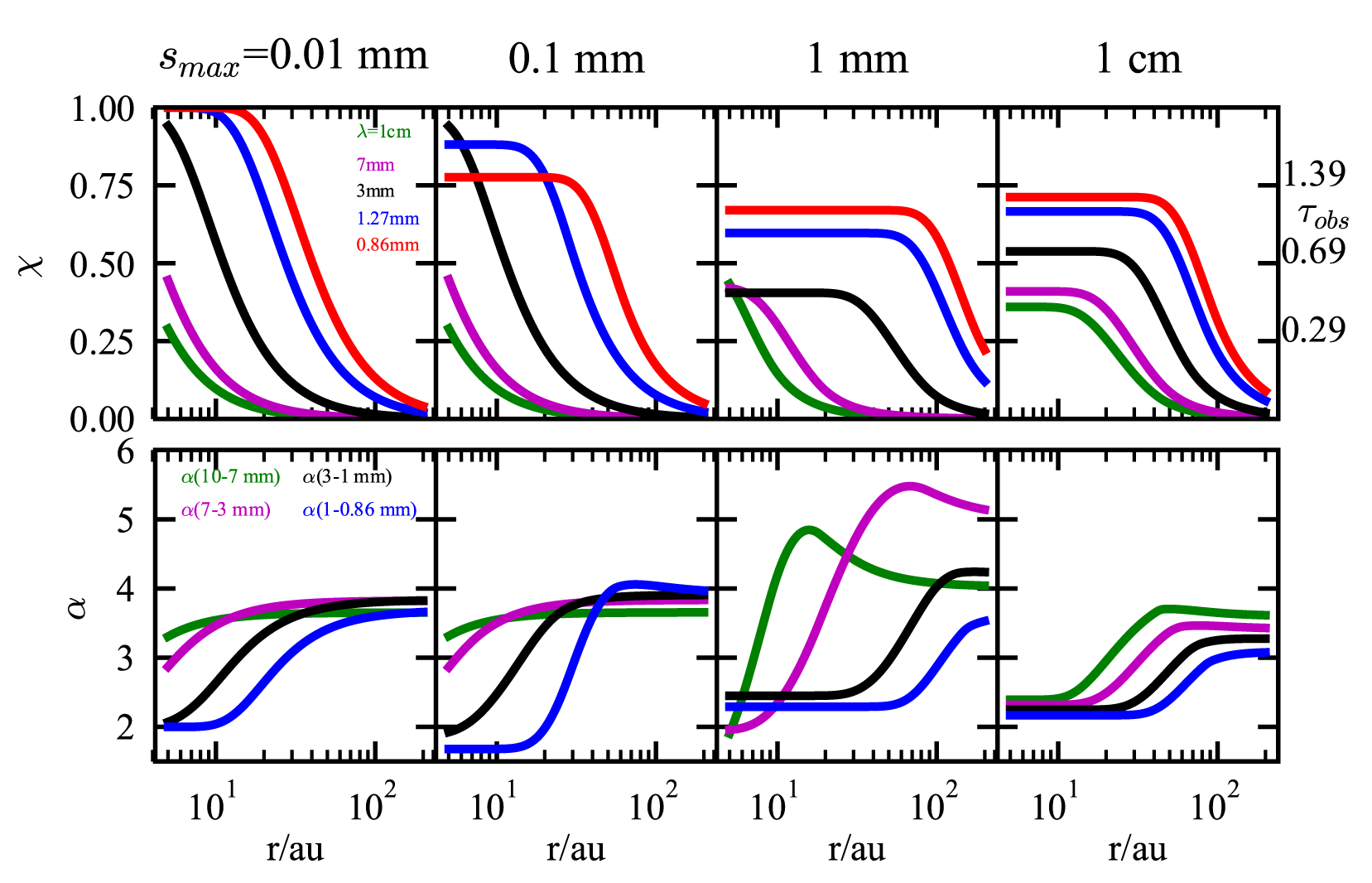}
\vspace{-4 mm}
\figcaption{{ $\chi$ and the spectral index $\alpha$ along the disk radius for a Q=1 disk with
different dust populations ($n(s)\propto s^{-3.5}$ for all the cases but $s_{max}$ increasing from left to right panels). 
At a particular wavelength, when the disk becomes optically thick within some radius, the $\chi$ curve flattens out. 
Similarly, if $\alpha$ becomes a constant $\sim$2 within some radius, the disk becomes optically thick at both of the wavelengths
that are used to measure $\alpha$. }\label{fig:radialvar}
}
\end{figure*}

\subsection{Dust Properties and the Spectral Index $\alpha$}

Just based on the DSHARP opacity, we want to explore how we can constrain the dust size distributions using
 the measured $\chi$ and the spectral index $\alpha$.
 The spectral index $\alpha$ is:
 \begin{equation}
\alpha\equiv\frac{{\rm ln}\left(\frac{I_{\nu_1}}{I_{\nu_2}}\right)}{{\rm ln}\left(\frac{\nu_{1}}{\nu_{2}}\right)}=\begin{cases} \frac{{\rm ln}\left(\frac{\chi_{\nu_1}B_{\nu_1}}{\chi_{\nu_2}B_{\nu_2}}\right)}{{\rm ln}\left(\frac{\nu_{1}}{\nu_{2}}\right)} \sim \frac{{\rm ln}\left(\frac{\chi_{\nu_1}}{\chi_{\nu_2}}\right)}{{\rm ln}\left(\frac{\nu_{1}}{\nu_{2}}\right)}+2 \,,\,\,\text{if}\,\,\text{thick} \\
  \frac{{\rm ln}\left(\frac{\kappa_{abs,\nu_1}B_{\nu_1}}{\kappa_{abs,\nu_2}B_{\nu_2}}\right)}{{\rm ln}\left(\frac{\nu_{1}}{\nu_{2}}\right)}\sim \frac{{\rm ln}\left(\frac{\kappa_{abs,\nu_1}}{\kappa_{abs,\nu_2}}\right)}{{\rm ln}\left(\frac{\nu_{1}}{\nu_{2}}\right)}+2 \,,\,\,\text{if}\,\,\text{thin} \,,\\
 \end{cases}
 \end{equation}
 where the approximation on the right is with the Rayleigh-Jeans approximation.
 Clearly, the measured $\alpha$ depends on totally different quantities in the optically thick and thin regimes. 
 In the more familiar optically thin regime, it depends on how the absorption coefficient changes with the frequency. 
 With $\kappa_{abs}\propto\nu^{\beta}$, $\alpha$ is $\beta+2$.
 In the optically thick regime, $\alpha$ depends on $\chi$. Assuming that $\chi\propto\nu^{\gamma}$, $\alpha$ is $\gamma+2$.
 As shown in Figure \ref{fig:omega}, $\omega$ changes monotonically with 1/$\chi$ that changes as $\lambda^{\gamma}$.
 Thus, if the albedo $\omega$ increases with wavelength ($\gamma>0$), $\alpha$=$\gamma$+2 measured at this wavelength span will be larger than 2, and vice versa.
To be more specific, if the disk is optically thick, we can use the measured $\alpha$ from observations to derive
 $\gamma$. Then we can constrain the relationship between $\omega_{\nu_1}$ and $\omega_{\nu_2}$ using 
 \begin{equation}
    \left(\frac{\nu_1}{\nu_2}\right)^\gamma=\frac{2\mu+\frac{\sqrt{3}}{\sqrt{1-\omega_{\nu_2}}}}{2\mu+\frac{\sqrt{3}}{\sqrt{1-\omega_{\nu_1}}}}\,, \label{eq:Inuoutthickratio}
\end{equation}
which is based on Equation \ref{eq:Inuoutthick2}.
  
Figure \ref{fig:smcombined} shows how $\kappa_{abs}$, $\chi$ and the spectral index $\alpha$ change with different dust populations
for different ALMA and VLA bands, using the DSHARP opacity. { The panels in the second row show $\chi$ in the optically thick limit.}
The third row shows $\alpha$ in the optically thin limit (thus denoted as $\alpha_{thin}$), while the bottom row is $\alpha$ in the optically thick limit ($\alpha_{thick}$). 
{ Here, we simply use $\beta+2$ or $\gamma+2$ to calculate $\alpha$, and thus do not consider the deviation of the blackbody radiation from the Rayleigh-Jeans limit \citep{huang18}.}
$\chi$ and both $\alpha$s have sharp transitions around $s_{max}\sim\lambda/2\pi$. When $s\ll\lambda/2\pi$, scattering is not important
and $\chi$=1, $\alpha_{thick}$=2, $\alpha_{thin}\sim 4$. When $s\gtrsim\lambda/2\pi$, $\chi$ becomes less than 1, $\alpha_{thick}$ becomes larger than 2, and
$\alpha_{thin}$ becomes less than 4. 
When the dust population has a lot of small dust (left panels), $\chi$ behaves more like a step function.  
When the disk is more populated with bigger particles (moving to the right panels), $\chi$ and both $\alpha$s have sharper peaks around
the transition. The peak of $\alpha_{thick}$ is due to the rapid  change of $\omega$ with wavelength for particles at bigger sizes. 
If $\omega$ increases with wavelength, $\chi$ will decrease with wavelength
and $\alpha_{thick}$ measured at this wavelength span will be larger than 2, and vice versa. { For example, for the $n(s)\propto s^{-3.5}$ 
and $s_{max}=1$ mm case in Figure \ref{fig:lambda}, $\omega$ increases with $\lambda$ at $\lambda<$ 4 mm and decreases with $\lambda$ at $\lambda>$ 4 mm,  
so the measured $\alpha_{thick}$ should be larger than 2 using observational bands 
$<$4 mm and smaller
than 2 using observational bands $>$4 mm. This is shown in Figure \ref{fig:smcombined}.
At $s_{max}=1$ mm, all $\alpha_{thick}$ curves for the $n(s)\propto s^{-3.5}$ case  have values larger than 2, except the green curve for 10-7 mm. }

Thus, in this optically thick disk with scattering scenario, $\alpha$ can be lower than 2,
which is not likely to happen for optically thin disks (e.g. the panels in the third row of Figure \ref{fig:smcombined}). If observations have measured that $\alpha$ is less than 2, it could
be a strong indication that the disk is optically thick and dust scattering plays an important role, having albedo decreasing with wavelength. 
Using the DSHARP opacity, $\alpha_{thick}$ panels in Figure \ref{fig:smcombined} suggest that, if the observed $\alpha$ from ALMA bands at $\lambda\lesssim$ 3 mm is less than 2, $s_{max}$
is  from 30 to 300 $\mu$m. When $s_{max}$ is larger than $\lambda/2\pi$, the probed $\omega$ is on the left side of the $\omega$ peak in Figure \ref{fig:lambda} so that $\alpha$ 
becomes larger than 2. This transition is relatively quick. After an $\alpha$ peak around $s_{max}\sim 0.5 \lambda$, $\alpha$ plateaus around a value slightly larger than 2. 

Consider a disk is optically thick at the inner disk and optically thin at the outer disk, the spectral index will be around 2 at the inner disk and suddenly change to 3-4 when $\tau<1$. This is simply because $\alpha$ in the optically thick and thin regimes are determined by different physical mechanisms. { To illustrate this point, 
we calculate $I_{\nu}/B_{\nu}$ and $\alpha$ for a Q=1 disk (Equation \ref{eq:Q1}) at both ALMA and VLA wavelength bands (Figure \ref{fig:radialvar}). At the inner disk where 
the disk is optically thick, $\chi$ and $\alpha$ flatten out. More disk region becomes optically thin with observations at longer wavelengths.   At VLA bands of 7 mm and 1 cm, 
the disk is optically thin even down to 5 au as long as $s_{max}\lesssim$ 1 mm. With $s_{max}=$1 cm which has the maximum opacity at $\sim$ 1 cm, VLA observations
can still probe the disk down to $20$ au before the disk becomes optically thick. 
} HL Tau observations by \cite{Carrasco2016} seem to indicate that $\alpha$ derived from ALMA bands
changes from 2 to 3 relatively quickly from 40 to 60 au, which may indicate that HL Tau is optically thick within 40 au in ALMA observations. On the other hand,
$\alpha$ derived from 3 mm ALMA and 7 mm VLA bands is larger than 2 throughout the disk, which indicates that the disk may be optically thin for VLA observations. 
Figure \ref{fig:radialvar} also suggests that we may want to measure the spectral index using every combination of two different bands, since $\alpha$ can change dramatically at different wavelengths,
especially at wavelengths close to the Mie resonances of the dust opacity.

{ Note that each panel in Figure \ref{fig:radialvar} assumes the same dust size distribution throughout the disk. In reality, dust size distributions vary both radially and vertically.
When the disk is optically thick, the emission is determined by the $\tau_{\nu}\sim1$ surface. So $\chi$ and $\alpha$ measured at the inner optically thick disk only inform us the dust
size distribution at the disk surface. The dust at the midplane could be a lot larger, which can only be probed by observations at longer wavelengths. When the disk is optically thin,
the measured $\alpha$ informs us the dust size distribution at the midplane directly (more discussion in Section 3.5 and Figure \ref{fig:scheme}).}

Previous spatially resolved $\alpha$ measurements have shown that $\alpha$ decreases towards the inner disk \citep{perez12, lperez15b, tazzari16, Carrasco2016}.
But $\alpha$ is always larger than 2 in these observations. TW Hya has some indications that $\alpha$ can be smaller than 2  at inner 20 au \citep{tsukagoshi16, huang18}. 
Although this deviation from 2 can in part be explained by the fact that the Rayleigh-Jeans approximation deviates from the blackbody radiation there \citep{huang18}, we discuss
the possibility that the disk has strong scattering and is optically thick at the inner disk. Under the optically thick scattering disk scenario, $\alpha<2$ within 20 au
implies that we have a large population of dust with $\sim$100 $\mu$m sizes there at the $\tau_{\nu}\sim$1 surface (based on the blue curves in the bottom row of Figure \ref{fig:smcombined}). 
At 25 au, observations suggest that there is an $\alpha$ peak
reaching $\alpha$=2.5, which is similar to the $\alpha_{thick}$ peak in Figure \ref{fig:smcombined}. Thus, one explanation for this peak is that dust size increases
with radius or we are probing deeper large-dust layers in the disk since the disk starts to become optically thin, 
and at 25 au there is a large population of $\sim$300 $\mu$m dust leading to the $\alpha_{thick}$ peak. 
Further out at 30 au, the dust becomes even bigger and $\alpha_{thick}$ decreases and plateaus, which is similar to the observations.  The increase of $\alpha$
at the outer disk may be due to the change in dust size or the whole disk becoming optically thin. 

Although this story may be too complicated, it indeed highlights that dust scattering in optically thick disks can also lead to gaps and rings.
ALMA continuum observations have revealed many gaps and rings in protoplanetary disks. Although some of these features are very prominent, some of them are very weak with only $\sim$20\% fluctuations (e.g. \citealt{huang18b}). These shallow features may also be explained by optically thick disks with radially varying dust scattering properties. If the dust becomes more reflective at a particular distance from the star, the disk will look like it has a gap there.  If the dust becomes less reflective there, the disk will look like it has a ring. 
Since changing the scattering properties means changing the dust compositions or distributions at the $\tau\sim$1 surface, the spectral index should also change at these gaps/rings. The spectral index can be either higher or lower at the rings, depending on how the particle size changes (e.g. Figure \ref{fig:smcombined}). On the other hand, we caution that dust scattering is unlikely 
to explain deep gaps (more than a factor of 10 deep) observed in some systems (e.g.
AS 209 \citealt{guzman18}), since a factor of 10 intensity reduction requires the albedo of 0.998 (based on Equation \ref{eq:omeganucal}) which is extremely high.

\subsection{Dust Disk Size - Luminosity Relationship}
Recent surveys \citep{tripathi17, andrews18a} confirm the previous hinted linear relationship \citep{andrews10b} between the submm continuum luminosity  and its emitting surface area. 
\cite{andrews18a} also confirm the disk luminosity and stellar luminosity relationship found in \cite{andrews13, ansdell16, pascucci16}.
\cite{andrews18a}  explore the optically thick disk scenario and find that the shapes of both relationships can be reproduced under this scenario. However, the optically thick
disks generate too much emission. To reduce the luminosities of optically thick disks to be consistent with observations, a filling factor of 0.3 in the disk is needed.
\cite{tripathi17} and \cite{andrews18a} suggest that substructures (e.g. rings, gaps) can lead to this filling factor.
Here, we suggest that, besides substructures, dust scattering can also decrease the luminosity for optically thick disks. Instead of the filling factor, a high albedo 
at the $\tau_{\nu}\sim1$ surface may also explain the observations.

\subsection{Connections with Previous Works}
Reducing blackbody intensity due to scattering is known in various astronomical communities (e.g. electron scattering reduces the radiation from accretion disks around compact objects).
For the protoplanetary disk study, \cite{miyake93} solve the radiative transfer equation for an isothermal disk (as summarized in \S 2.1). These results have been mentioned in many works afterwards (e.g. \citealt{dalessio01, birnstiel18}). \cite{sierra17}  applies this to dusty vortices and show that the optically thick vortex center becomes fainter if dust scattering is considered. 

\begin{figure}[t!]
\includegraphics[width=3.4in]{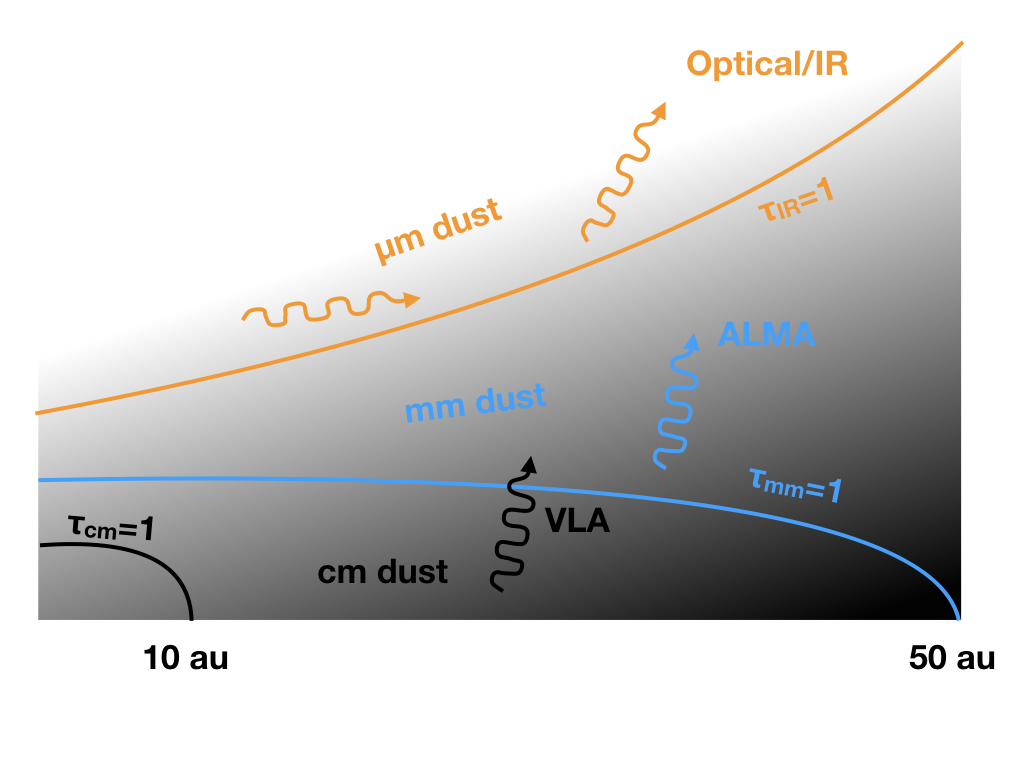}
\vspace{-10 mm}
\figcaption{{ The schematic diagram showing the dust distribution in a protoplanetary disk and $\tau=1$ surface for observations at different wavelengths. 
The disk within $\sim$50 au can be optically thick to
ALMA short millimeter wavelength observations, so that these observations actually probe mm dust slightly above the midplane. VLA observations probe the dust
at the midplane which can be larger than mm.   }\label{fig:scheme}
} 
\end{figure}

On the other hand, dust scattering is largely ignored in radio observations since the protoplanetary disk is thought to be optically thin so that dust scattering is not important. Recent works by \cite{kataoka15} and \cite{yang16a} have suggested that dust scattering may be crucial for explaining submm polarization measurements, although other mechanisms may still be needed to explain the observations \citep{kataoka17, yang16b}. However, there is a strong tension between the dust size constrained by polarization measurements and submm-cm continuum spectral index measurements \citep{kataoka16}. { Here, we suggest that such tension may be due to the optically thin assumption in both polarization and submm continuum studies. If the disk is optically thick, the spectral index is normally smaller than that from an optically thin disk.  Assuming that the disk is optically thin, submm continuum observations can overestimate the particle size significantly. In reality (the schematic diagram from Figure \ref{fig:scheme}), the small $\alpha$
by submm observations could simply reflect the disk is optically thick for these observations and dust at the $\tau_{mm}\sim 1$ surface has a typical size of 0.1-1 mm with strong scattering. 
For longer wavelength
observations by VLA \citep{lperez15b}, the disk is likely to be optically thin beyond 10 au. The small $\alpha$ measured from these observations could indeed imply big particles at deeper layers  (likely the midplane) in the disk where VLA is probing, which is a natural outcome from dust settling. Note that the absorption opacity at 7 mm is 10-50 times smaller than the opacity at 1.25 mm. Thus, VLA
probes a much deeper layer in the disk. MCRT calculations for disks with such vertically varied dust-size-distribution will be presented in Zhang et al. (In prep).}

{ Another line of evidence that most protoplanetary disks are optically thick at ALMA bands is that inclined disks are systematically less massive than face-on disks based on $\sim$1 mm observations assuming that the disks are optically thin (Figure 4 in \citealt{garufi18}). If the disks are optically thin, the measured dust mass should not depend on the disk inclination. 
Thus, this inclination dependence suggests that the disks are optically thick at ALMA bands.
Furthermore, with scattering included, we suggest that inclined optically thick disks
will look even fainter (Figure \ref{fig:mu}). This effect may explain the very low temperature derived from the edge-on disk Flying Saucer \citep{guilloteau16}.}

Although the optically thick scattering disk scenario seems to be promising to explain several observations, there is evidence that the disks are not optically thick everywhere. 
The first is that the rings beyond 40 au in the DSHARP sample can be well fitted with Gaussian profiles along the radial direction, 
instead of flat-topped profiles \citep{dullemond18}. Since the distribution of dust trapped in turbulent disks with rings should
follows a Gaussian profile, the intensity profile will have a flat top instead of a Gaussian profile if these rings are optically thick at the ring center. The second line of evidence against the optically thick rings is that
CO emission coming from the back side of HD 163296 is dimmer at the location of the bright dust rings at 67 and 100 au \citep{isella18}. This dimming is due to dust extinction by the rings. 
Since the CO emission does not disappear completely, the dust rings cannot be optically thick. However, these two lines of 
evidence only apply to the
rings at the outer disk beyond 40 au. This is actually consistent with our Figure \ref{fig:scat}, where even a Q=1 disk will be optically thin beyond 50 au. 
The optically thick scattering disk scenario only applies to the inner disk within 50 au. We need to carry out similar tests or use other methods (e.g. \citealt{harsono18}, \citealt{powell17}) to study the optical depth and mass of the inner disk in future.

{ One assumption in this work is that the disk is isothermal in the vertical direction. This is not quite correct since the dusty disk intercepts the stellar irradiation causing
the temperature inversion at the disk atmosphere \citep{calvet91}.
On the other hand, large dust in protoplanetary disks settles to the disk midplane (probed by radio observations) while small dust is still suspended at the disk atmosphere (probed by near-IR observations). 
Small dust intercepts the stellar irradiation which determines the disk temperature structure. Small dust radiates energy vertically towards the disk midplane to warm up the large dust.
Since large dust sits in the thermal bath generated by small dust, we expect that it should be approximately isothermal. The MCRT simulations for such disk configuration will be presented in Zhang et al. (in prep). Furthermore, the emission reduction argument in the abstract and our preliminary
simulations show that, if the disk is optically thick but not isothermal, the emission reduction still applies and
is mainly determined by the dust scattering properties at $\tau_{\nu}\sim1$. 
Thus, if the disk having large dust
is optically thick but not isothermal, the emission reduction probes the albedo of the $\tau_{\nu}\sim1$ surface. }

\section{Conclusion \label{sec:conclusion}}
ALMA protostar surveys have 
suggested that the dust (with sizes of $\lesssim$ cm) in Class II protostellar disks may not be enough to explain the averaged solid mass in exoplanets, leading
to the speculation that a large fraction of dust mass has already been converted to planetesimals at the Class II stage. 
On the other hand, the dust mass derivation from ALMA observations is based on the assumption that 
protostellar disks are optically thin at submm.
This optically thin assumption seems to be supported by 
recent high angular resolution observations from the DSHARP ALMA survey where
the measured optical depths of most DSHARP disks are less than one.

However, in this work, we point out that dust scattering is important for the disk mass estimate, at least within the inner 50 au.
Using the analytical theory, direct numerical simulations, and MCRT calculations, we have shown that 
dust scattering can reduce the emission from an optically thick region.  
Ignoring dust scattering can lead to an underestimate of the disk optical depth, and an optically thick disk with 
dust scattering can be misidentified as an optically thin disk. When the disk is more inclined, optically thick scattering makes the disk look even fainter.
{ When the disk is large (e.g. 100 au),
most of dust mass is at the outer disk which is optically thin at 1.25 mm, so that 1.25 mm observations will only underestimate the disk mass by a factor of $\sim$2 for these extended disks. 
When the disk is compact ($<$ 30 au), ALMA 1.25 mm observations can easily underestimate the real dust mass by a factor of 10. On the other hand, for VLA 7 mm observations,
the disk is optically thin beyond several au.  So VLA observations provide a much more accurate mass estimate.}

When the disk is optically thick, we can measure $\chi$ or $\tau_{obs}$ if we know the 
disk temperature and use them to constrain the dust albedo following the well-defined
simple relationships  (Equation \ref{eq:Inuoutthick}, \ref{eq:Inuoutthick2}, and \ref{eq:omeganucal}).
The measured optical depth of 0.6 in the DSHARP disks can be naturally explained by optically thick dust with an albedo of $\sim$0.9 at 1.25 mm (Equation \ref{eq:omeganucal}).
Using the DSHARP opacity, this albedo corresponds to a dust population with the maximum grain size of 0.1-1 mm.

If we have multi-band observations, we can also use the spectral index $\alpha$ to constrain dust properties.
In the optically thick regime, the spectral index $\alpha$ depends on the albedo $\omega$ rather than $\kappa$ as in the optically thin regime.
If $\omega$ increases with wavelength, $\alpha$ measured at this wavelength span will be larger than 2, and vice versa. 
Using the DSHARP opacity, if the observed $\alpha$ from ALMA is less than 2, the dust
is smaller than  $\sim$300 $\mu$m, vice versa. We also find that
$\alpha$ is normally smaller than 2.5 in the optically thick regime and larger than 2.5 in the optically thin regime. Thus, we expect to see a jump of $\alpha$ when the disk
changes from optically thick to optically thin along the radial direction. 
We discuss the possibility that radial changes in $\alpha$ in TW Hya might be related to the change
of dust properties along the radial direction. 

This optically thick disk scenario also provides an explanation for the known submm luminosity-disk size relationship, 
and may ease the strong tension between the dust size constrained by polarization measurements and 
submm-cm continuum spectral index measurements. { The small $\alpha$ and high polarization degree from submm observatoins
could be due to that the disk is optically thick for these observations and dust at the $\tau_{mm}\sim1$ surface has a typical size
of 0.1-1 mm with strong scattering. For VLA observations at longer wavelengths, the disk is optically thin and the small $\alpha$
measured from these observations could imply big particles at the deeper layer or the disk midplane, which is a natural outcome from dust settling. } 

We suggests that dust in protoplanetary disks may be hidden from ALMA observations at short millimeter wavelengths, and
longer wavelength observations (e.g. ngVLA and SKA) are desired. Properly modeling
dust continuum emission including dust scattering is crucial for constraining disk structures. 
Optically thick disks with scattering also
provide unique opportunities to study dust properties in protoplanetary disks.

\section*{Acknowledgements}
Z. Z. thank the organizers of "Planet-Forming Disks, a workshop to honor Antonella Natta" to organize the great workshop, during which the $\tau\sim0.6$ problem was raised many times and the idea of this work came to mind. The Planet Forming Disks workshop has received funding from the European UnionÕs Horizon 2020 research and innovation programme under grant agreement No 730562 [RadioNet], No 743029 [EASY], from INAF-Arcetri and iALMA contract No 6041, DIGDEEP contract No 706320. Z. Z. thanks 
Carlos Carrasco-Gonz{\'a}lez,  Anibal Sierra, Nuria Calvet, and Lee Hartmann for comments on the initial draft. 
Z. Z. acknowledge Ian Rabago for proofreading the initial draft. Z. Z. also thank Francois M{\'e}nard for interesting discussion on highly inclined disks. 
Z. Z. acknowledges support from the National Science Foundation under CAREER Grant Number AST-1753168
and Sloan Research Fellowship. 
T.B. acknowledges funding from the European Research Council (ERC) under the European Union's Horizon 2020 research and innovation programme under grant agreement No~714769.
C.P.D. acknowledges support by the German Science Foundation (DFG) Research Unit FOR~2634, grants DU~414/22-1 and DU~414/23-1.
S. A. and J. H. acknowledge funding support from the National Aeronautics and Space Administration under grant No.~17-XRP17$\_$2-0012 issued through the Exoplanets Research Program.
J.H. acknowledges support from the National Science Foundation Graduate Research Fellowship under Grant No. DGE-1144152.
L.P. acknowledges support from CONICYT project Basal AFB-170002 and from FCFM/U. de Chile Fondo de Instalaci\'on Acad\'emica.
Simulations are carried out with the support from the Texas Advanced Computing Center (TACC)
at The University of Texas Austin through XSEDE grant TG-AST130002, and 
the NASA High-End Computing (HEC) Program through the NASA Advanced Supercomputing (NAS)
Division at Ames Research Center.

\clearpage

\end{document}